\documentclass[preprint,preprintnumbers,amsmath,amssymb]{revtex4}
\usepackage{graphicx}% Include figure files
\usepackage{dcolumn}% Align table columns on decimal point
\usepackage{bm}% bold math

\begin{document}

%\preprint{APS/123-QED}

\title{On the onset of interference effects during the formation of the Bose-Einstein condensate}% Force line breaks with \\

\author{Miguel Escobedo}
 \altaffiliation[Electronic address: ]{miguel.escobedo@ehu.es}%Lines break automatically or can be forced with \\
\affiliation{%
Departamento de Matem\'aticas\\ Universidad del Pa{\'\i}s Vasco\\
Apartado 644 Bilbao 48080 Spain.
}%

\author{J. J. L. Vel\'azquez}
  \altaffiliation[Electronic address: ]{JJ\_Velazquez@mat.ucm.es}
\affiliation{
Departamento de Matem\'atica Aplicada\\
Facultad de Matem\'aticas\\
Universidad Complutense \\
Madrid 28040 Spain.
}%

\date{\today}% It is always \today, today,
             %  but any date may be explicitly specified
\begin{abstract}

In this paper we derive the equations characterizing the boundary
layer which describes the transition of the distribution function of a gas
of weakly interacting bosons to the distribution function of the gas in the
presence of a Bose-Einstein condensate. To this end, we first rederive the
classical Uehling-Uhlenbeck equation taking as a starting point the dynamics
of a system of many weakly interacting quantum particles. The solutions of
the Uehling-Uhlenbeck equation yield blow-up in finite time. Near the
blow-up time the approximations used to derive the Uehling-Uhlenbeck
equation break down. We derive the set of equations that describe the
building of correlations and the onset of quantum interference effects for
the many-particle hamiltonian system under the assumption that the blow-up
for the Uehling-Uhlenbeck equation takes place in a self-similar form.

\end{abstract}

%\pacs{Valid PACS appear here}% PACS, the Physics and Astronomy
                             % Classification Scheme.
%\keywords{Suggested keywords}%Use showkeys class option if keyword
                              %display desired
\maketitle

\section{\label{sec:1}Introduction.}

The kinetic equation that describes the evolution of a rarefied system of
bosons was obtained in \cite{10} by L. W. Nordheim and in \cite{14} by E. A.
Uehling and G. E. Uhlenbeck. In the limit of Born's
approximation,  the resulting equation, also known as quantum
Boltzmann equation is the following:
\begin{align}
&\frac{\partial f}{\partial t}   +\frac{p_1}{m}\nabla_{x}
f(p_1,x,t)=C(f,f)\label{S0E1}\\
&C(f,f)   =\frac{4 \pi  g^2}{\hbar }\int  
\frac{dp_{2}}{(2\pi\hbar)^3}\int \frac{dp_{3}}{(2\pi\hbar)^3}\int
\frac{dp_{4}}{(2\pi\hbar)^3}\, (2\pi\hbar)^3 
\delta(\varepsilon (p_1)+\varepsilon (p_{2})-\varepsilon (p_{3})
-\varepsilon (p_{4}))\times \nonumber\\
&\hskip 11.5cm \times q[f](p_1, p_2; p_3, p_4, t) \label{S0E2}
\\ 
&q[f] (p_1, p_2; p_3, p_4, t)=\left[f(p_3) f(p_4)\!\left(1+f(p_1)\right)\!\left(1+f(p_2)\right)-\right. \nonumber \\
&\left.\hskip 9.1cm - f(p_1)f(p_2)\!\left(1+\!f(p_3)\right)\!\left(1+f(p_4)\right)\right]\label{S0E3}
\end{align}
with $g=4\pi a\hbar^{2}/m$ where $m$ is the mass of the particles,
 $\varepsilon (p)=|p|^2/2m$  is the energy of the particles and $a$ is
the s-wave scattering length  (cf. \cite{15} for example). 

The starting point for the derivation of a kinetic system of
equations is a set of equations describing the dynamics of a
system of $N$ particles. In the case of weakly interacting bosons
such dynamics is described by the Schr\"{o}dinger equation for a
system of $N$ interacting particles. Assuming that the interaction
between different particles are weak enough it is possible to
obtain suitable evolution equations for the one-particle
distribution function using a perturbative method. For classical
particles, this has been mathematically proved for short times
(cf. \cite{20})  or globally in time for special situations  (cf.
\cite{21}). This is the standard method used to derive the
Uehling-Uhlenbeck equation (see for example the classical
monographies \cite{16, 2},   as well as  the more mathematically
oriented approaches in \cite{18}, \cite{23}, \cite{ESY},  and  \cite{22}). 
Similar arguments for the Fermionic case may be found in \cite{Hu} and \cite{HL}.

It turns out that the solutions of
(\ref{S0E1})-(\ref{S0E3}) can develop singularities in finite
time, as it has been obtained in the numerical
simulations for spatially homogeneous distributions of particles in
\cite{11}, \cite{12} as well as in \cite{9}. The interpretation of this blow up phenomenon,  given by the authors of these papers, is that such event corresponds to the formation of the B-E condensate. As we will see in this paper, the derivation of the U-U equation, taking as a starting point a quantum many particles system, is not valid near the time of the formation of  the singularity.

On the other hand, the quantum dynamics of the particles in the condensate is described by the Gross Pitaevskii equation (cf. \cite{3, 4, 13, 15}). A rigorous mathematical proof of the precise formulation of this fact has been obtained in \cite{25} for the stationary case and in \cite{24} for the non equilibrium case with short range interactions in suitable scaling limits.

We are interested in the process formation of the condensate, a question which has already been considered by several authors, see for example  {\cite{BS}, \cite{KS}, \cite{13}.}
Our main goal is to  describe in a detailed manner
how the transition between the kinetic regime described by the U-U equation, and the quantum regime described by means of a nonlinear Schr\"odinger equation.
The relevant non-dimensional
parameter is the quotient $\Delta E\, \Delta t /\hbar$, where $\Delta E$ is a characteristic value of the energy and $\Delta T$ is a characteristic time scale for the variation of the density distributions. Interference effects cannot be
ignored if this non-dimensional number becomes of order one. We
derive then the equation of the boundary layer which should
describe in detail the transition from the distribution function
at the critical time to the density function in which the
condensate is present.
\\ \\
Part of the arguments presented in this paper are standard in statistical physics. For example we make extensive use of the second quantification formalism in Section II. We also use the classical BBGKY hierarchy for quantum particles in Section III. The arguments used in these two Sections are also used in the theory of weak turbulence and in general in the derivation of kinetic equations for weakly interacting waves (cf. \cite{BZ}, \cite{Z}). Part of the arguments of Section IV are reminiscent of similar ones in \cite{ESY} and \cite{15}. The main result of the paper is contained in Sections V and VI. We have however included the results in the previous Sections in order to fix the notation and explain the limit under consideration.

\section{\label{sec:2}The N-particles system and the Second quantification formalism .}

We start recalling the classical derivation of the U-U equation that takes
as starting point the study of the dynamics of a quantum
$N-$particle system by means of the second quantification
formalism. This will allow us to precise the assumptions in which
such a derivation is based.

 Let us assume that we have $N$ quantum particles contained in a box
$\Omega\equiv\left[ 0,L\right] ^{3}.$ We will denote the density
of particles as $n=\frac{N}{L^{3}}.$ We will suppose also that the
particles interact by means of pair potentials. The hamiltonian of
the system is then given by:
\begin{equation}
H_{N}=H_{0,N}+H_{1,N} \label{S1ham}%
\end{equation}
where:
\begin{equation}
H_{0,N}=-\frac{\hbar^{2}}{2m}\sum_{j=1}^{N}\Delta_{x_{j}},\;H_{1,N}=\frac{1}{2}\sum
_{k=1}^{N}\sum_{j=1,\;k\neq j}^{N}V\left(  x_{k}-x_{j}\right)  .
\label{S1ham1}%
\end{equation}
The evolution of the system is given by means of Heisenberg's equation for the
density matrix:
\begin{align}
i\hbar\partial_{t}\rho &  =\left[  H_{N},\rho\right] \label{S1Ea}\\
\rho\left(  0\right)   &  =\rho_{0}. \label{S1Eb}%
\end{align}
The precise form of the initial density matrix that characterizes
the initial state of the system will be given
later. Let us
precise the order of magnitude of the several parameters arising
in the system. There are two main characteristic length scales in the
problem, namely the average distance between particles
$d=\frac{L}{ N^{1/3}},$ and the De Broglie length that is given by
$\lambda=\frac{\hbar }{p_{0}}\equiv\frac{\hbar}{\sqrt{2mk_{B}T}}.$
By assumption $p_{0}\equiv \sqrt{2mk_{B}T}$ is just an estimate of
a typical particle momentum. The temperature $T$ is not a true
thermodynamic temperature, because the system is not in
equilibrium, but it is just a measure of the characteristic energy
for the gas particles.

On the other hand, Born's approximation means in mathematical
terms that we may assume  that the interaction potential between
particles  $V$ is smooth, but after deriving a set of kinetic
equations we will take the limit $V\rightarrow g\,
\delta\left(x\right)$, where $g$ is defined just after the formula
(\ref{S0E3}).

The main assumptions on the physical parameters that we use in this paper are the
following:
\begin{align}
& N    >>1,\;\;L>>1,\;\;n=\frac{N}{L^{3}}=\text{constant}\label{S1E1}\\
& \lambda   \sim d \label{S1E2}\\
& \frac{m\lambda ^2 g}{d^3}\sim \frac{\lambda ^2 a}{d^3}<<1.
\label{S1E3}%
\end{align}
Assumption (\ref{S1E1}) is the usual Thermodynamic limit assumption that ensures that there are not boundary effects on the resulting equations. Assumption
(\ref{S1E3}) means that the particle interactions are weak and allows us to derive a kinetic equation for the evolution of the distribution of particles. Finally assumption  (\ref{S1E2}) means that the particle densities are large enough to allow for the formation of B-E condensate. This is related to the fact that  the kinetic equation obtained under the assumption (\ref{S1E2}) can yield blow up in finite time

We will first obtain a set of kinetic equations describing the
evolution of the solutions of (\ref{S1Ea}), (\ref{S1Eb}) in the
limit defined by (\ref{S1E1})-(\ref{S1E3}). This problem was
solved by Nordheim and Uehling \& Uhlenbeck under the implicit
additional hypothesis of the boundedness  for the solution of the
resulting kinetic equation. However, such assumption
fails, because the solution of the limit equation blows up in
finite time as it has been seen in the numerical simulations of
\cite{9}, \cite{11, 12}. Therefore the Uehling-Uhlenbeck equation
is not the correct limit for the
system of particles under consideration in the limit (\ref{S1E1}%
)-(\ref{S1E3}), if the time $t$ is close to the time of formation
of a singularity. The main
goal of this paper is to obtain a new kinetic equation describing
the distribution of particles during the formation of
the condensate.

\subsection{\label{1}Second quantification formalism.}

In order to study the $N$ particle system in the limit (\ref{S1E1}%
)-(\ref{S1E3}) we will use the formalism of the second
quantification. Most of the computations in this Subsection are
standard, but we will reproduce them for the reader's convenience.
We will assume that the hamiltonian $H_{N}$ in (\ref{S1ham}),
(\ref{S1ham1}) acts in the Hilbert
space:
\begin{equation}
\mathcal{H}_{N}\equiv\bigotimes_{n=0}^{N}\left(  L^{2}\left(  \Omega\right)
\right)  ^{n} \label{S1E6}%
\end{equation}

By definiteness we will assume that the wave functions satisfy periodic
boundary conditions in $\Omega.$ Homogeneous Dirichlet boundary conditions
would work similarly. For periodic boundary conditions the
eigenvalues of the momentum operator for a single
particle  $p_{k}
\equiv-i\hbar\partial_{x_{k}}$ are given by:
\[
p=\frac{2\pi\hbar}{L}\ell,\;\;\ell\in\mathbb{Z}^{3}%
\]

We will denote the normalized eigenfunctions associated to the operator
$H_{0,N}$ as:
\[
\left|  ...,n_{\ell},....\right\rangle \;\;\;,\;\;\ell\in\mathbb{Z}^{3}%
\]
where $n_{\ell}$ is the number of particles in the state $\ell.$

For notational convenience we will use also the following
alternative way of writing these eigenfunctions:
\[
\left|  ...,n_{\ell},....\right\rangle=\left|  n\left(  \ell\right)  \right\rangle ,\;\text{where\ }n:\mathbb{Z}%
^{3}\rightarrow\mathbb{N}=\left\{  0,1,2,...\right\}
\]
This notation will be convenient to write in a short manner all
the possible choices of occupation numbers.

We introduce the well known  annihilation and creation operators
$a_{\ell},\;a_{\ell}^{+}$, $N_\ell$ whose action on these eigenfunctions is
given by:
\begin{align}
a_{\ell}\left|  ...,n_{\ell},....\right\rangle  &  =\sqrt{n_{\ell}}\left|
...,n_{\ell}-1,....\right\rangle \;\;\;\;,\;\;\ell\in\mathbb{Z}^{3}%
\label{S1E6a}\\
a_{\ell}^{+}\left|  ...,n_{\ell},....\right\rangle  &  =\sqrt{n_{\ell}%
+1}\left|  ...,n_{\ell}+1,....\right\rangle \;\;\;\;,\;\;\ell\in\mathbb{Z}^{3}
\label{S1E6b}\\
N_{\ell}\left|  ...,n_{\ell},....\right\rangle &\equiv a_{\ell}^{+}a_{\ell
}\left|  ...,n_{\ell},....\right\rangle =n_{\ell}\left|  ...,n_{\ell
},....\right\rangle
\end{align}
These operators satisfy the commutation relations:%
\begin{align}
\label{S1E7}
\left[  a_{k},a_{\ell}^{+}\right]    =\delta_{k,\ell},\quad \left[  a_{k},a_{\ell}\right]   &  =\left[  a_{k}^{+},a_{\ell}^{+}\right]
=0.
\end{align}
We now define the annihilation  and creation operators of a particle at the point $x$ of
$\Omega$ by means of:
\begin{equation}
\psi\left(  x\right)  =\frac{1}{L^{\frac{3}{2}}}\sum_{\ell\in\mathbb{Z}^{3}%
}a_{\ell}e^{\frac{2\pi i\ell x}{L}},\;\psi^{+}\left(  x\right)  =\frac
{1}{L^{\frac{3}{2}}}\sum_{\ell\in\mathbb{Z}^{3}}a_{\ell}^{+}e^{-\frac{2\pi
i\ell x}{L}} \label{S1E8}%
\end{equation}
Notice that using (\ref{S1E7}):
\begin{align*}
\left[  \psi\left(  x\right)  ,\psi^{+}\left(  y\right)  \right]  & =\frac{1}{L^{3}}\sum_{\ell\in\mathbb{Z}^{3}}e^{\frac{2\pi i\ell\left(
x-y\right)  }{L}}=\delta\left(  x-y\right) \\
\left[  \psi\left(  x\right)  ,\psi\left(  y\right)  \right]   &  =\left[
\psi^{+}\left(  x\right)  ,\psi^{+}\left(  y\right)  \right]  =0
\end{align*}
Using all these operators we can rewrite the operator $H_{0,N}$ as:
\[
H_{0,N}=\sum_{j=1}^{N}\frac{p_{j}^{2}}{2m}=\sum_{\ell\in\mathbb{Z}^{3}%
}\epsilon_{\ell}a_{\ell}^{+}a_{\ell}%
\]
where:
\[
\epsilon_{\ell}\equiv\frac{4\pi^{2}\hbar^{2}\ell^{2}}{2mL^{2}}\;\;\;,\;\;\ell
\in\mathbb{Z}^{3}%
\]

Taking the gradient of (\ref{S1E8}) we obtain: whence:
\begin{equation}
H_{0,N}=\frac{\hbar^{2}}{2m}\int_{\Omega}\nabla\psi^{+}\left(  x\right)
\nabla\psi\left(  x\right)  dx \label{S1E9}%
\end{equation}

On the other hand:
\begin{align}
H_{1,N}   =\frac{1}{2}\int_{\Omega}dx_{1}\int_{\Omega}dx_{2}V\left(
x_{1}-x_{2}\right)\psi^{+}\left(  x_{1}\right)  \psi^{+}\left(  x_{2}\right)
\psi\left(  x_{1}\right)  \psi\left(  x_{2}\right)  \label{S1E10}%
\end{align}
We define the  distribution functions :
\begin{align}
 f_{j,m}\left(  x_{1},...,x_{j}; y_{1},...,y_{m}\right)  \equiv
 \operatorname*{Tr}\left(  \rho\psi^{+}\left(  y_{1}\right)  \psi^{+}\left(
y_{2}\right)  ...\psi^{+}\left(  y_{m}\right)  \psi\left(  x_{1}\right)
\psi\left(  x_{2}\right)  ...\psi\left(  x_{j}\right)  \right) \label{S1E11}
\end{align}
The computation of the evolution equations for the functions
$f_{j,m}$ is standard (see for example \cite{16}) . Using (\ref{S1Ea}) it follows that:
\begin{align}
i\hbar\partial_{t}f_{j,m} =\operatorname*{Tr}\left(  \rho\left[  \psi
^{+}\left(  y_{1}\right)  \psi^{+}\left(  y_{2}\right)  ...\psi^{+}\left(
y_{m}\right)\psi\left(  x_{1}\right)  \psi\left(  x_{2}\right)
...\psi\left(  x_{j}\right)  ,H_{N}\right]  \right)\label{S1ECom}
\end{align}

On the other hand, we can compute the commutator in (\ref{S1ECom})
to obtain the following evolution equation for the distribution
functions $f_{j,m}:$
\begin{align}
&i\hbar\partial_{t}f_{j,m}\left( x_{1},...,x_{j};y_{1},...,y_{m}\right) =-
\frac{\hbar^{2}}{2m}\left(
\sum_{s=1}^{j}\Delta_{x_{s}}-\sum_{s=1}^{m}\Delta_{y_{s}}\right)
f_{j,m}\left( x_{1},...,x_{j};y_{1},...,y_{m}\right) + \notag\\
& +\int_{\Omega}d\xi\left[ \sum_{k=1}^{j}V\left( \xi-x_{k}\right)
-\sum_{k=1}^{m}V\left( \xi-y_{k}\right) \right] f_{j+1,m+1}\left(
x_{1},...,x_{j},\xi;y_{1},...,y_{m},\xi\right) +  \notag\\
& +\frac{1}{2}\left[ \sum_{k=1}^{j}\sum_{s=1\;,\;k\neq s}^{j}V\left( x_{k}-x_{s}\right)
-\sum_{k=1}^{m}\sum_{s=1}^{m}V\left( y_{k}-y_{s}\right) \right] f_{j,m}%
\left( x_{1},...,x_{j};y_{1},...,y_{m}\right) .   \label{S2E1}
\end{align}

\subsection{\label{2}On the choice of the initial data.}

In order to solve the system of equations (\ref{S2E1}) we must
prescribe suitable initial data. We will assume that the initial
matrix density $\rho\left(  0\right)$ satisfies:

%??It is not the macrocanonical distribution that corresponds to
%equilibrium. ????
\begin{align}
\rho\left(  0\right)   &  =\rho_{0}=\frac{1}{Q}\sum_{n:\mathbb{Z}%
^{3}\rightarrow N}P_{0}\left(  z,\Theta ;n\right)  \left|  n\right\rangle
\left\langle n\right| \label{S2E5}\\
P_{0}\left(  z,\Theta ;n\right)   &  \equiv z^{\left[  \sum_{\ell\in\mathbb{Z}^{3}%
}n(\ell)\right]  }\left(  \prod_{\ell\in\mathbb{Z}^{3}}
(\theta_{\ell})^{n(\ell)}\right)
\label{S2E6}\\
n  &  :\mathbb{Z}^{3}\rightarrow\mathbb{N}\;\;,\;\;\sum_{\ell\in\mathbb{Z}%
^{3}}n\left(  \ell\right)  <\infty\label{S2E7}%
\end{align}
where:
\[
Q=\sum_{n:\mathbb{Z}^{3}\rightarrow N}P_{0}\left(  z,\Theta ;n\right)
\]
has been chosen in order to have $\operatorname*{Tr}\left(
\rho_{0}\right) =1,$ and where:
\[
\theta_{\ell}\equiv \Theta \left(  \frac{2\pi^{2}\hbar^{2}}{mk_{B}T}
\frac{\ell^{2}}{L^{2}}\right)  .
\]
Choosing the initial data as in (\ref{S2E5}), (\ref{S2E6}) we are
assuming that the particles are independently and homogeneously
distributed in space according to the distribution
$\Theta \left(\cdot\right)$ in the space of energy. Since we use a  macrocanonical distribution the number of variables ia stochastic variable. In the thermodynamic limit the
fluctuations in the number of particles can be expected to
disappear as it is usual in statistical physics.
The value of $z$ is chosen  to obtain a given  average
number of particles $N$ for the distribution. Therefore:

\begin{align*}
\left\langle N\right\rangle =\operatorname*{Tr}\left(  \rho_{0}N\right)  
&  =\frac{1}{Q}\sum_{n:\mathbb{Z}^{3}\rightarrow N}P_{0}\left(  z,\Theta ;n\right)
N\left(  n\right)=z\frac{\partial\left(  \log\left(  Q\right)  \right)  }{\partial
z}\end{align*}
where $  N\left(  n\right)  =\sum_{\ell\in\mathbb{Z}^{3}}n\left(
\ell\right)$.

Instead of analyzing the original system (\ref{S1Ea}),
(\ref{S1Eb}), we will study the equivalent system of equations
(\ref{S2E1}) that is more convenient to use perturbative
arguments. Due to (\ref{S1E11}) we must solve these equations with
initial data:
\begin{align}
&  f_{j,m;0}\left(  x_{1},...,x_{j};y_{1},...,y_{m}\right)  \equiv
\label{S2E8}\\
&  \operatorname*{Tr}\left(  \rho_{0}\psi^{+}\left(  x_{1}\right)  \psi
^{+}\left(  x_{2}\right)  ...\psi^{+}\left(  x_{j}\right)  \psi\left(
y_{1}\right)  \psi\left(  y_{2}\right)  ...\psi\left(  y_{m}\right)  \right)
.\nonumber
\end{align}

Using (\ref{S2E5}), (\ref{S2E6}) we obtain:
\begin{equation}
f_{j,m;0}\left(  x_{1},...,x_{j};y_{1},...,y_{m}\right)  =0\;\;\text{if\ \ }%
j\neq m \label{S2E9}%
\end{equation}

The evolution equations (\ref{S2E1}), with initial data
(\ref{S2E9}) admit a solution
satisfying:
\[
f_{j,m}\left(  x_{1},...,x_{j};y_{1},...,y_{m};t\right)  =0\;\;\text{if\ \ }%
j\neq m
\]

Therefore, we can restrict our study to the functions:
\begin{equation*}
F_{k}\left(  x_{1},...,x_{k};y_{1},...,y_{k};t\right)  \equiv  f_{k,k}\left(
x_{1},...,x_{k};y_{1},...,y_{k};t\right)  \label{S2E10}%
\end{equation*}

On the other hand we can compute the initial distribution \hfill
\break$F_{k,0}\left( x_{1},...,x_{k};y_{1},...,y_{k}\right) \equiv
F_{k}\left( x_{1},...,x_{k};y_{1},...,y_{k};0\right)$ that due
to (\ref{S1E8}) and (\ref{S2E8}) is given by:

\begin{align}
F_{k,0}\left( x_{1},...,x_{k};y_{1},...,y_{k}\right)
 = &\frac{1}{L^{3k}}\sum_{\ell_{1}\in\mathbb{Z}^{3}}...\sum_{\ell_{k}\in
\mathbb{Z}^{3}}\sum_{j_{1}\in\mathbb{Z}^{3}}...\sum_{j_{k}\in\mathbb{Z}^{3}%
}e^{-\frac{2\pi i\left(
\ell_{1}x_{1}+...+\ell_{k}x_{k}\right) }{L}+\frac{2\pi i\left(
j_{1}y_{1}+...+j_{k}y_{k}\right) }{L}}\times \nonumber \\
 & \times \operatorname*{Tr}\left(
\rho_{0}a_{\ell_{1}}^{+}a_{\ell_{2}}^{+}...a_{%
\ell_{k}}^{+}a_{j_{1}}...a_{j_{k}}\right).\label{S2E12}
\end{align}

By (\ref{S2E5}) we then have:
\begin{align*}
\operatorname*{Tr}\left(  \rho_{0}a_{\ell_{1}}^{+}a_{\ell_{2}}^{+}%
...a_{\ell_{k}}^{+}a_{j_{1}}...a_{j_{k}}\right)  = \frac{1}{Q}\sum
_{n:\mathbb{Z}^{3}\rightarrow N}\prod_{\ell\in\mathbb{Z}^{3}}\left(
z\theta_{\ell}\right)  ^{n\left(  \ell\right)  }\left\langle n\right|  a_{\ell_{1}}^{+}a_{\ell_{2}}^{+}...a_{\ell_{k}%
}^{+}a_{j_{1}}...a_{j_{k}}\left|  n\right\rangle
\end{align*}
We now compute the terms $\left\langle n\right|  a_{\ell_{1}}^{+}a_{\ell_{2}%
}^{+}...a_{\ell_{k}}^{+}a_{j_{1}}...a_{j_{k}}\left|  n\right\rangle .$ We will
assume for the moment that all the terms $j_{1},...,j_{k}$ are different. In
this case, the matrix element $\left\langle n\right|  a_{\ell_{1}}^{+}%
a_{\ell_{2}}^{+}...a_{\ell_{k}}^{+}a_{j_{1}}...a_{j_{k}}\left|  n\right\rangle
$ is not zero if and only if the coefficients $\ell_{1},...,\ell_{k}$ are the
same as the $j_{1},...,j_{k}$ or a permutation of them. Therefore:
\begin{align}
\left\langle n\right|  a_{\ell_{1}}^{+}a_{\ell_{2}}^{+}...a_{\ell_{k}}%
^{+}a_{j_{1}}...a_{j_{k}}\left|  n\right\rangle =\left[  \prod_{s=1}%
^{k}n\left(  j_{s}\right)  \right]  \sum_{\sigma\in S^{k}}\delta_{j_{\sigma\left(  1\right)  },\ell_{1}%
}\delta_{j_{\sigma\left(  2\right)  },\ell_{2}}...\delta_{j_{\sigma\left(
k\right)  },\ell_{k}} \label{S2E13}%
\end{align}
where $S^{k}$ is the group of permutations of the elements $\left\{
1,...,k\right\}  .$ If some of the coefficients $j_{1},...,j_{k}$ are repeated,
the scalar product on the left hand side of (\ref{S2E13})
can be bounded, using standard statistical physics computations (cf. \cite{2}),  as $\frac{C}{L^3}\left[  \prod_{s=1}^{k}n\left(j_{s}\right)  \right]$. Therefore, using formula (\ref{S2E13}) again and replacing some sums by Riemann integrals, we can approximate (\ref{S2E12}) in the limit $L\to \infty$ as
\begin{equation}
F_{k,0}\left(  x_{1},...,x_{k};y_{1},...,y_{k}\right)  =\sum_{\sigma\in S^{k}%
}\prod_{m=1}^{k}F_{0}\left(  y_{m}-x_{\sigma\left(  m\right)  };z\right)
\label{S2E17}%
\end{equation}

where:
\begin{align}
&F_{0}\left(  y;z\right)  \equiv\int_{\mathbb{R}^{3}}\left[  \frac{z \Theta\left(\overline \epsilon(\xi)\right)  e^{2\pi iy\xi}}{1-z \Theta \left(
\overline \epsilon(\xi)\right)  }\right]  d\xi ,\qquad \overline \epsilon (\xi)\equiv\frac{2\pi^{2}\hbar^{2}}{mk_{B}T}\xi^{2}. \label{S2E16b}
\end{align}

\section{\label{sec:3} The small correlations approximation.}

\subsection{Non dimensional equations.}

Summarizing, we have reduced the problem to the following system of
equations (cf. (\ref{S2E1}), (\ref{S2E17})):
\begin{align}
&  i\hbar\partial_{t}F_{k}\left(  x_{1},...,x_{k};y_{1},...,y_{k}, t\right)
=A_{1}+A_{2}+A_{3}; \qquad k=1,2, \cdots  \label{S3E1}\\
&  A_{1}=- \frac{\hbar^{2}}{2m}\left(  \sum_{s=1}^{k}\left[  \Delta_{x_{s}%
}-\Delta_{y_{s}} \right]  \right)  F_{k}\left(  x_{1},...,x_{k};y_{1}%
,...,y_{k}; t\right) \label{S3E1-100}\\
&  A_{2}=\int_{\Omega}d\xi\left[  \sum_{j=1}^{k}\left[  V\left(  \xi
-x_{j}\right)  -V\left(  \xi-y_{j}\right)  \right]  \right]F_{k+1}\left(  x_{1},...,x_{k},\xi;y_{1},...,y_{k},\xi; t\right) \label{S3E1-200}\\
&  A_{3}=\frac{1}{2}\left[  \sum_{j=1}^{k}\sum_{s=1\;,\;j\neq s}^{k}\left[  V\left(
x_{j}-x_{s}\right)  -V\left(  y_{j}-y_{s}\right)  \right]  \right]
 F_{k}\left(  x_{1},...,x_{k};y_{1},...,y_{k}; t\right) \label{S3E1-300}
\end{align}

with initial data:
\begin{equation}
F_{k,0}\left(  x_{1},...,x_{k};y_{1},...,y_{k}\right)  =\sum_{\sigma\in S^{k}%
}\prod_{m=1}^{k}F_{0}\left(  y_{m}-x_{\sigma\left(  m\right)  };z\right).
\label{S3E2}%
\end{equation}
The two first equations of this hierarchy are:
\begin{align}
 &i\hbar\partial_{t}F_{1}\left(  x_{1};y_{1};t\right)  =-\frac{\hbar^{2}}%
{2m}\left(  \Delta_{x_{1}}-\Delta_{y_{1}}\right)  F_{1}\left(  x_{1}%
;y_{1};t\right) \nonumber \\
&+\int_{\Omega}d\xi\left[  V\left(  \xi-x_{1}\right)  -V\left(  \xi
-y_{1}\right)  \right]  \left[  F_{1}\left(  x_{1};y_{1}; t\right)  F_{1}\left(
\xi;\xi; t\right)+F_{1}\left(  x_{1};\xi; t\right)  F_{1}\left(  \xi;y_{1}; t\right)
\right] \nonumber \\
&\hskip 2cm +\int_{\Omega}d\xi\left[  V\left(  \xi-x_{1}\right)  -V\left(  \xi
-y_{1}\right)  \right]  F_{2}\left(  x_{1},\xi;y_{1},\xi; t\right)  \label{S3E5}
\end{align}

\begin{align}
 i\hbar\partial_{t}F_{2}  \left(  x_{1},x_{2};y_{1},y_{2}; t\right)
= -\frac{\hbar^{2}}{2m}\left(  \Delta_{x_{1}}+\Delta_{x_{2}}  -\Delta_{y_{1}%
}-\Delta_{y_{2}}\right)  F_{2}\left(  x_{1},x_{2};y_{1},y_{2}; t\right)
+\nonumber\\
 +\sum_{j=1}^{2}\int_{\Omega}\!\!d\xi\left[  V\left(  \xi-x_{j}\right)
\!\!-\!\!V\left(  \xi-y_{j}\right)  \right] { F}_{3}\left(  x_{1},x_{2},\xi;y_{1},y_{2},\xi; t\right)
+\nonumber\\
 +\left[  V\left(  x_{1}-x_{2}\right)  -V\left(  y_{1}-y_{2}\right)
\right]  { F}_{2}\left(  x_{1}, x_{2}; y_{1}, y_{2}; t\right)
\label{S3E6}
\end{align}

In order to understand more clearly the limit that we are considering, we introduce the non-dimensional variables:
\begin{align}
x & =  \lambda\, \hat x, \quad 
V(x)=\frac{g}{\lambda ^3}\,\widehat V(\hat x), \quad
F_k(x)=\left(\frac{1}{d^3} \right)^k\, \widehat F _k(\hat x),\quad t= \frac{2m\lambda^2}{\hbar\, \epsilon^2 }\, \hat t, \quad
p=\frac{\hbar}{\lambda}\, \hat p
\label{S3E600000}\\
 g & =  \epsilon\frac{\hbar ^2}{2 m\lambda^2}d^3  \label{S3E600001}
\end{align}
where, due to (\ref{S1E3}), $\epsilon$ is a small parameter and by assumption, the potential $\widehat V (\hat x)$ is now of order one. Our choice of time scale is due to the fact that we want to obtain, in the limit $\epsilon \to 0$,  an equation in which the particle density varies in times $\hat t $ of order one. Notice that $(x, p)\to (\hat x, \hat p)$ is not a canonical transformation but transforms a ``quantum cell'' in the phase space of volume $\hbar$ into another cell of volume one.

For the sake of simplicity we drop the hats in all the variables $\hat x$, $\hat p$, $\hat t$, $\widehat F_k$ and  $\widehat V$ as it is customary in the computations of asymptotic expansions. The system (\ref{S3E5})-(\ref{S3E6}) then becomes

\begin{align}
 i\partial_{t}F_{1}\left(  x_{1};y_{1};t\right)  =-\frac{1}%
{\epsilon^2}\left(  \Delta_{x_{1}}-\Delta_{y_{1}}\right) & F_{1}\left(  x_{1}%
;y_{1};t\right) \nonumber \\
+\frac{1}%
{\epsilon}\int_{\Omega}d\xi\left[  V\left(  \xi-x_{1}\right)  -V\left(  \xi
-y_{1}\right)  \right] & \left[  F_{1}\left(  x_{1};y_{1}; t\right)  F_{1}\left(
\xi;\xi; t\right)+F_{1}\left(  x_{1};\xi; t\right)  F_{1}\left(  \xi;y_{1}; t\right)
\right] \nonumber \\
&+\frac{1}%
{\epsilon}\int_{\Omega}d\xi\left[  V\left(  \xi-x_{1}\right)  -V\left(  \xi
-y_{1}\right)  \right]  F_{2}\left(  x_{1},\xi;y_{1},\xi; t\right)  \label{S3E5TER}
\end{align}

\begin{align}
 i\partial_{t}F_{2}  \left(  x_{1},x_{2};y_{1},y_{2}; t\right)
= -\frac{1}%
{\epsilon^2}\left(  \Delta_{x_{1}}+\Delta_{x_{2}}  -\Delta_{y_{1}%
}-\Delta_{y_{2}}\right)  F_{2}\left(  x_{1},x_{2};y_{1},y_{2}; t\right)
+\nonumber\\
 +\frac{1}%
{\epsilon}\sum_{j=1}^{2}\int_{\Omega}\!\!d\xi\left[  V\left(  \xi-x_{j}\right)
\!\!-\!\!V\left(  \xi-y_{j}\right)  \right] {F}_{3}\left(  x_{1},x_{2},\xi;y_{1},y_{2},\xi; t\right)
+\nonumber\\
 +\frac{1}{\epsilon}\left(\frac{d}{\lambda }\right)^3\left[  V\left(  x_{1}-x_{2}\right)  -V\left(  y_{1}-y_{2}\right)
\right]  { F}_{2}\left(  x_{1}, x_{2}; y_{1}, y_{2}; t\right)
\label{S3E6TER}
\end{align}

\subsection{Small correlations limit.}

Our aim is to obtain closure relations for the functions $F_{k}$
by means of a perturbative  argument. 

Notice that in
the absence of potential the system of equations (\ref{S3E5TER}),
(\ref{S3E6TER}) might be explicitly solved and the resulting
solutions have the form:
\begin{equation}
F_{k}\left(  x_{1},...,x_{k};y_{1},...,y_{k};t\right)  =\sum_{\sigma\in S^{k}%
}\prod_{m=1}^{k}F_{1}\left(  x_{\sigma\left(  m\right)  },y_{m};t\right)
\label{S3E3}%
\end{equation}

Moreover, the function $F_{0}$ in (\ref{S2E16b}) is invariant under spatial translations whence $F_{0}\left(  x_{1}%
;y_{1}\right)  =F_{0}\left(  x_{1}-y_{1}\right)$. Since the system of 
equations (\ref{S3E1})-(\ref{S3E1-300}) are also invariant under spatial
translations it follows that $F_{1}\left( x_{1};y_{1};t\right)
=F_{1}\left( x_{1}-y_{1};t\right)  $ for any $t>0.$

Notice that in this case we can think in the solutions of this
form as ``uncorrelated'' solutions, although in a strict
mathematical sense the corresponding probability distributions are
not uncorrelated, but the only correlations between particles
whose distribution is given by (\ref{S3E3}) would be the ones due
to the symmetry of the wave functions due to the bosonic  character of the particles (cf. the
discussion in \cite{2}). In any case the approximation
(\ref{S3E3}) is a convenient starting point for the computation of
the solutions of (\ref{S3E1})-(\ref{S3E2}) in a perturbative
manner. We define the correlation functions $G_k$ by means of the identity:
\begin{align}
&  F_{k}\left(  x_{1},...,x_{k};y_{1},...,y_{k};t\right)  =G_{k}\left(
x_{1},...,x_{k};y_{1},...,y_{k};t\right)  +\nonumber\\
&  + {\widetilde F}_{k}\left(  x_{1},...,x_{k};y_{1},...,y_{k};t\right).
\label{S3E4}
\end{align}
where we have defined
\begin{align}
&  {\widetilde F}_{k}\left(  x_{1},...,x_{k};y_{1},...,y_{k};t\right)
\!\!=\!\!\sum_{\sigma\in S^{k}}\!\prod_{m=1}^{k}\!\!F_{1}\left(
x_{\sigma\left(  m\right)  },y_{m};t\right) \label{S3E4bis}.
\end{align}
It is possible to derive a kinetic approximation for (\ref{S3E5}) (\ref{S3E6}) under the following small correlation assumptions

\begin{equation}
\label{S3E4bis0090}
|G_k|<<\prod_{m=1}^{k}|F_1|.
\end{equation}

Indeed, under this assumption we obtain, plugging (\ref{S3E4}) into (\ref{S3E5})-(\ref{S3E6}):
\begin{align}
 i\partial_{t}F_{1}\left(  x_{1};y_{1};t\right)  =-\frac{1}%
{\epsilon^2}\left(  \Delta_{x_{1}}-\Delta_{y_{1}}\right) & F_{1}\left(  x_{1}%
;y_{1};t\right) \nonumber \\
+\frac{1}%
{\epsilon}\int_{\Omega}d\xi\left[  V\left(  \xi-x_{1}\right)  -V\left(  \xi
-y_{1}\right)  \right] & \left[  F_{1}\left(  x_{1};y_{1}; t\right)  F_{1}\left(
\xi;\xi; t\right)+F_{1}\left(  x_{1};\xi; t\right)  F_{1}\left(  \xi;y_{1}; t\right)
\right] \nonumber \\
&+\frac{1}%
{\epsilon}\int_{\Omega}d\xi\left[  V\left(  \xi-x_{1}\right)  -V\left(  \xi
-y_{1}\right)  \right]  F_{2}\left(  x_{1},\xi;y_{1},\xi; t\right)  \label{S3E5BIS}
\end{align}

\begin{align}
 i\partial_{t}F_{2}  \left(  x_{1},x_{2};y_{1},y_{2}; t\right)
= -\frac{1}%
{\epsilon^2}\left(  \Delta_{x_{1}}+\Delta_{x_{2}}  -\Delta_{y_{1}%
}-\Delta_{y_{2}}\right)  F_{2}\left(  x_{1},x_{2};y_{1},y_{2}; t\right)
+\nonumber\\
 +\frac{1}%
{\epsilon}\sum_{j=1}^{2}\int_{\Omega}\!\!d\xi\left[  V\left(  \xi-x_{j}\right)
\!\!-\!\!V\left(  \xi-y_{j}\right)  \right] {\widetilde F}_{3}\left(  x_{1},x_{2},\xi;y_{1},y_{2},\xi; t\right)
+\nonumber\\
 +\frac{1}{\epsilon}\left(\frac{d}{\lambda }\right)^3\left[  V\left(  x_{1}-x_{2}\right)  -V\left(  y_{1}-y_{2}\right)
\right]  { \widetilde F}_{2}\left(  x_{1}, x_{2}; y_{1}, y_{2}; t\right)
\label{S3E6BIS}
\end{align}
The relative strength of the terms yielding correlations is of order  $\epsilon$. This explains why in equation (\ref{S3E6BIS}) we have approximated $F_2$ and $F_3$ by $\widetilde F_2$ and $\widetilde F_3$ respectively. Notice that we have kept all the terms in the equation (\ref{S3E5BIS}) and
only terms of order $1/\epsilon$ or larger in (\ref{S3E6BIS}). 

We now compute the evolution equation for $G_{2}\left(  x_{1},x_{2};y_{1},y_{2}; t\right)  $ using (\ref{S3E4}) and the approximation (\ref{S3E5BIS})-(\ref{S3E6BIS}): 
\begin{eqnarray}
i\partial_{t}G_{2}\left( x_{1},x_{2};y_{1},y_{2}; t\right) =
-\frac{1}{\epsilon^2}\left(
\Delta_{x_{1}}+\Delta_{x_{2}}-\Delta_{y_{1}}-\Delta_{y_{2}}\right)
G_{2}\left( x_{1},x_{2};y_{1},y_{2}; t\right) +\notag\\
+\frac{1}{\epsilon}\left(\frac{d}{\lambda }\right)^3\left[ V\left( x_{1}-x_{2}\right) -V\left( y_{1}-y_{2}\right) \right] %
{\widetilde F}_{2}\left( x_1, x_2; y_1, y_2; t\right)+
\frac{1}{\epsilon}\int_{\Omega}d\xi
H(\xi; t) \label{S3E9a}
\\ \nonumber \\
  H(\xi; t)=\sum_{j=1}^{2} \left[  V\left(  \xi-x_{j}\right)  -V\left(
\xi-y_{j}\right)  \right] {\widetilde F}_{3}\left(  x_{1},x_{2},\xi;y_{1},y_{2},\xi; t\right) \nonumber \\
-\sum_{\sigma\in S^{2}}\left[  V\left(  \xi-x_{1}\right)  -V\left(
\xi-y_{\sigma\left(  1\right)  }\right)  \right]  F_{2}\left(  x_{1},\xi;y_{\sigma\left(  1\right)  },\xi; t\right)
F_{1}\left(  x_{2};y_{\sigma\left(  2\right)  }; t\right) \nonumber\\
 -\sum_{\sigma\in S^{2}}F_{1}\left(  x_{1};y_{\sigma\left(  1\right)
}; t\right)  \left[  V\left(  \xi-x_{2}\right)  -V\left(  \xi-y_{\sigma\left(
2\right)  }\right)  \right] F_{2}\left(  x_{2},\xi;y_{\sigma\left(  2\right)  },\xi; t\right)
\end{eqnarray}
After some computations, neglecting terms of order ${\cal O}(\epsilon ^2)$ it follows that:
\begin{align}
&H(\xi; t)    = V\left(  \xi-x_{1}\right)  F_{1}\left(  x_{2};\xi; t\right)
{\widetilde F}_{2}\left(  x_{1},\xi;y_{1},y_{2}; t\right)  - V\left(  \xi-y_{1}\right)  F_{1}\left(  \xi;y_{2}; t\right)  {\widetilde
F}_{2}\left(  x_{1},x_{2};y_{1},\xi; t\right) \nonumber\\
&  + V\left(  \xi-x_{2}\right)  F_{1}\left(  x_{1};\xi; t\right)  {\widetilde
F}_{2}\left(  x_{2},\xi;y_{1},y_{2}; t\right)  - V\left(  \xi-y_{2}\right)  F_{1}\left(  \xi;y_{1}; t\right)  {\widetilde
F}_{2}\left(  x_{1},x_{2};y_{2},\xi; t\right).  \label{S3E11}%
\end{align}
Plugging (\ref{S3E11}) into (\ref{S3E9a}) we obtain:
\begin{align}
&  i\partial_{t}G_{2}\left(  x_{1},x_{2};y_{1},y_{2}; t\right)
=-\frac{1}{\epsilon^2}\left(  \Delta_{x_{1}}+\Delta_{x_{2}}-\Delta_{y_{1}%
}-\Delta_{y_{2}}\right)  G_{2}\left(  x_{1},x_{2};y_{1},y_{2}; t\right)
\label{S3E12}\\
&  +\frac{1}{\epsilon}\int\left[  V\left(  \xi-x_{1}\right)  F_{1}\left(  x_{2};\xi; t\right)
{\widetilde F}_{2}\left(  x_{1},\xi;y_{1},y_{2}; t\right) -V\left(  \xi-y_{1}\right)  F_{1}\left(  \xi;y_{2}; t\right)
{\widetilde F}_{2}\left(  x_{1},x_{2};y_{1},\xi; t\right)  \right]
d\xi\nonumber\\
& + \frac{1}{\epsilon}\int\left[  V\left(  \xi-x_{2}\right)  F_{1}\left(  x_{1};\xi; t\right)
{\widetilde F}_{2}\left(  x_{2},\xi;y_{1},y_{2}; t\right) -V\left(  \xi-y_{2}\right)  F_{1}\left(  \xi;y_{1}; t\right)
{\widetilde F}_{2}\left(  x_{1},x_{2};y_{2},\xi; t\right)  \right]
d\xi\nonumber\\
&\hskip 5cm  +
\frac{1}{\epsilon}\left(\frac{d}{\lambda }\right)^3\left[  V\left(  x_{1}-x_{2}\right)  -V\left(  y_{1}-y_{2}\right)
\right]  {\widetilde F}_{2}\left(  x_{1}, x_{2}; y_{1}, y_{2}; t\right),\nonumber
\end{align}
where, due to (\ref{S1E2}), $d/\lambda$ is of order one.\\ 
The equations (\ref{S3E5BIS}) and (\ref{S3E12}) provide the evolution equations for the functions $F_{1}$ and $G_{2}$. We have not used yet Born's approximation which in the variables that we are using reduces to
\begin{equation}
V\left(  x\right)  =\delta\left(  x\right)  \label{S3E13}
\end{equation}
Using (\ref{S3E13}) we obtain:
\begin{align}
  i\partial_{t}F_{1}\left(  x_{1};y_{1}; t\right)  =-\frac{1}%
{\epsilon ^2}&\left(  \Delta_{x_{1}}-\Delta_{y_{1}}\right)  F_{1}\left(  x_{1}%
;y_{1}; t\right)  +\nonumber\\
&  +\frac{1}%
{\epsilon }g\left[  G_{2}\left(  x_{1},x_{1};y_{1},x_{1}; t\right)
-G_{2}\left(  x_{1},y_{1};y_{1},y_{1}; t\right)  \right] \label{S3E14}
\end{align}
\begin{align}
 i\partial_{t}G_{2}\left(  x_{1},x_{2};y_{1},y_{2}; t\right)
=-\frac{1}{\epsilon^2}\left(  \Delta_{x_{1}}+\Delta_{x_{2}}-\Delta_{y_{1}%
}-\Delta_{y_{2}}\right)  G_{2}\left(  x_{1},x_{2};y_{1},y_{2}; t\right)
\nonumber\\
  +\frac{1}{\epsilon}g\left[  F_{1}\left(  x_{2};x_{1}; t\right)  {\widetilde F}%
_{2}\left(  x_{1},x_{1};y_{1},y_{2}; t\right)  -F_{1}\left(  y_{1};y_{2}; t\right)  {\widetilde F}_{2}\left(
x_{1},x_{2};y_{1},y_{1}; t\right)  \right]  +\nonumber\\
 +\frac{1}{\epsilon}g\left[  F_{1}\left(  x_{1};x_{2}; t\right)  {\widetilde F}%
_{2}\left(  x_{2},x_{2};y_{1},y_{2}; t\right)  -F_{1}\left(  y_{2};y_{1}; t\right)  {\widetilde F}_{2}\left(
x_{1},x_{2};y_{2},y_{2}; t\right)  \right] \nonumber\\
 +\frac{1}{\epsilon}\left(\frac{d}{\lambda }\right)^3g\left[  \delta\left(  x_{1}-x_{2}\right)  -\delta\left(
y_{1}-y_{2}\right)  \right]  {\widetilde F}_{2}\left(  x_{1},x_{2};y_{2}%
,y_{2}; t\right) \label{S3E15}
\end{align}

The invariance of the initial distribution $F_{0}\left(  x_{1};y_{1}\right)  $
under spatial translations imply that, with a slight abuse of language, the solutions of (\ref{S3E14}), (\ref{S3E15}) have the form:
\begin{align}
F_{1}\left(  x_{1};y_{1}; t\right)   &  = F_{1}\left(  x_{1}-y_{1}; t\right) \label{S3E15b} \\
G_{2}\left(  x_{1},x_{2};y_{1},y_{2}; t\right)   &  = G_{2}\left(  x_{1}%
-y_{1},x_{2}-y_{1};0,y_{2}-y_{1}; t\right)  \label{S3E15a}%
\end{align}

Under these assumptions the equations (\ref{S3E14}), (\ref{S3E15}) reduce to:

\begin{eqnarray}
&&i\partial_{t}F_{1}\left( x_{1}-y_{1}; t\right) =\frac{1}{\varepsilon}g\left[
G_{2}\left( x_{1},x_{1};y_{1},x_{1}; t\right) -G_{2}\left(
x_{1},y_{1};y_{1},y_{1}; t\right) \right]  \label{S3E16-1}\\
&& i\partial_{t}G_{2}\left( x_{1},x_{2};y_{1},y_{2}; t\right)
=-\frac{1}{\varepsilon^2}\left(
\Delta_{x_{1}}+\Delta_{x_{2}}-\Delta_{y_{1}}-\Delta_{y_{2}}\right)
G_{2}\left( x_{1},x_{2};y_{1},y_{2}; t\right) +\frac{1}{\varepsilon}\,  Q\left[ F_{1}\right]\nonumber \\
\label{S3E16}
\end{eqnarray}

where:

\begin{align}
 Q\left[  F_{1}\right]  \left(  x_{1},x_{2};y_{1},y_{2}; t\right)  =\left[
F_{1}\left(  x_{2};x_{1}; t\right)  {\widetilde F}_{2}\left(  x_{1},x_{1};y_{1},y_{2}; t\right)   -F_{1}\left(  y_{1};y_{2}; t\right)  {\widetilde F}_{2}\left(
x_{1},x_{2};y_{1},y_{1}; t\right)  \right]  \nonumber\\
  +\left[  F_{1}\left(  x_{1};x_{2}; t\right)  {\widetilde F}_{2}\left(
x_{2},x_{2};y_{1},y_{2}; t\right)  -F_{1}\left(  y_{2};y_{1}; t\right)  {\widetilde F}_{2}\left(
x_{1},x_{2};y_{2},y_{2}; t\right)  \right]  \nonumber\\
  +\left(\frac{d}{\lambda} \right)^3\left[  \delta\left(  x_{1}-x_{2}\right)  -\delta\left(  y_{1}%
-y_{2}\right)  \right]  {\widetilde F}_{2}\left(  x_{1},x_{2};y_{2}%
,y_{2}; t\right)  \label{S3E17}%
\end{align}

Due to (\ref{S3E2}) we have:
\begin{equation}
G_{2}\left(  x_{1}, x_{2};y_{1}, y_{2};0\right)  \equiv0. \label{S3E4bis1}%
\end{equation}

The system of equations (\ref{S3E16-1})-(\ref{S3E4bis1}) will be
our starting point for the description of the Bose gas in which we
are interested. Notice that it is a closed system of partial
differential equations.

\section{\label{sec:4} The problem in the phase space.}

The function that describes the one-particle density in the phase
space in quantum problems is the Wigner transform of $F_{1}\left(
x_{1};y_{1}; t\right) .$ Such a function is defined
as:
\begin{equation}
f_{1}\left(  x,p; t\right)  =\frac{1}{(2 \pi)^3}
\int_{\mathbb{R}^{3}}F_{1}\left(  x+\zeta; x-\zeta; t\right)
e^{i\zeta p}d\zeta\label{S4E1}%
\end{equation}
where the normalization constant in (\ref{S4E1}) has been
chosen in order to have
\begin{equation*}
\label{S4E1bis}
\int f_{1}\left(  x,p\right) dxdp=N.
\end{equation*}
In the spatially homogeneous case we have $F_{1}\left(  x+\zeta
;x-\zeta; t\right)  =F_{1}\left(  \zeta; t\right)  $ due to
(\ref{S3E15a}). Therefore, (\ref{S4E1}) reduces to the Fourier
transform:
\begin{equation}
f_{1}\left(  x,p; t\right)  =f_{1}\left(  p, t\right)  =\frac{1}{(2 \pi)^3}\int_{\mathbb{R}^{3}}F_{1}\left(  \zeta; t\right)
e^{i\zeta p}d\zeta\label{S4E2}%
\end{equation}

In order to obtain the evolution equation for $f_{1}\left(  p, t\right)  $ we
then take the Fourier transform of (\ref{S3E16}), (\ref{S3E17}):
\begin{align}
i\partial_{t}f_{1}\left(  p, t\right)  =\frac{1}{\left(
2\pi\right)  ^{3}\epsilon}\int_{\mathbb{R}^{3}}\left[  G_{2}\left(  \zeta
,\zeta;0,\zeta; t\right)  -  G_{2}\left(  \zeta,0;0,0; t\right)  \right]  e^{i\zeta p
}d\zeta\label{S4E3}%
\end{align}

On the other hand we have the following Fourier representation for the
function $G_{2}\left(  x_{1},x_{2};y_{1},y_{2}; t\right)  $:
\begin{align}
&  g_{2}\left(  \xi_{1},\xi_{2};\eta_{1},\eta_{2}; t\right)  =\frac{1}{\left(
2\pi\right)  ^{12}}\int_{\left(  \mathbb{R}^{3}\right)  ^{4}}dx_{1}%
dx_{2}dy_{1}dy_{2} e^{i\left(  \xi_{1}x_{1}+\xi_{2}x_{2}-\eta_{1}y_{1}-\eta
_{2}y_{2}\right)  }G_{2}\left(  x_{1},x_{2};y_{1},y_{2}; t\right) \label{S4E4}\\
&  G_{2}\left(  x_{1},x_{2};y_{1},y_{2}; t\right)
=\int_{\left(  \mathbb{R}%
^{3}\right)  ^{4}}d\xi_{1}d\xi_{2}d\eta_{1}d\eta_{2}e^{-i\left(  \xi_{1}x_{1}+\xi_{2}x_{2}-\eta_{1}y_{1}-\eta
_{2}y_{2}\right)  }g_{2}\left(  \xi_{1},\xi_{2};\eta_{1},\eta_{2}; t\right)  .
\label{S4E5a}%
\end{align}
Let us write:
\begin{align}
w\left(  \xi_{1},\xi_{2};\eta_{1},\eta_{2}; t\right)  =\frac{1}{\left(  2\pi
\right)  ^{12}}\int_{\left(  \mathbb{R}^{3}\right)  ^{4}}dx_{1}%
dx_{2}dy_{1}dy_{2}e^{i\left(  \xi_{1}x_{1}+\xi_{2}x_{2}-\eta_{1}y_{1}-\eta
_{2}y_{2}\right)  }Q\left[  F_{1}\right]  \left(  x_{1},x_{2};y_{1}%
,y_{2}; t\right)  \label{S4E5b}%
\end{align}

Taking the Fourier transform of (\ref{S3E16}) and
using (\ref{S4E5a}) in (\ref{S4E3}) we obtain the following system of
equations for $f_{1}$ and $g_{2}:$

\begin{eqnarray}
&&i\partial_{t}f_{1}\left( p, t\right)  =\frac{1}{\epsilon}\, \int_{\left(
\mathbb{R}^{3}\right) ^{4}}d\xi_{1}d\xi_{2}d\eta_{1}d\eta_{2}\left
[ \delta\left(
p-\eta_{1}\right) -\delta\left( p-\xi_{1}\right) \right] g_{2}\left(
\xi_{1},\xi_{2};\eta_{1},\eta_{2}; t\right)  \label{S4E9} \\
&&i\partial_{t}g_{2}\left( \xi_{1},\xi_{2};\eta_{1},\eta_{2}; t\right)  =%
\frac{1}{\epsilon^2}\left[ \varepsilon \left( \xi_{1}\right)+\varepsilon\left( \xi_{2}\right)
-\varepsilon\left( \eta_{1}\right)-\varepsilon\left( \eta_{2}\right)\right]
g_{2}\left( \xi_{1},\xi_{2};\eta_{1},\eta_{2}; t\right)+ \nonumber \\
&& \hskip 10cm+\frac{1}{\epsilon}\,  w\left(
\xi_{1},\xi_{2};\eta_{1},\eta_{2}; t\right)  \label{S4E10}
\end{eqnarray}

where $w\left(  \xi_{1},\xi_{2};\eta_{1},\eta_{2}; t \right)  $ is as in
(\ref{S3E17}), (\ref{S4E5b}) and the energy $\varepsilon(p)$ in the non-dimensional variables is $\varepsilon(p)=p^2$.
To obtain a closed system for $f_1, g_2$ it only remains to
compute $w\left( \xi_{1},\xi_{2};\eta_{1},\eta_{2}; t\right) $ in
terms of $f_1$. To this end notice that (\ref{S4E2}) yields:
\begin{equation}
F_{1}\left(  \zeta; t\right)  =\int f_{1}\left(  p, t\right)  e^{-ip\zeta
}dp
\end{equation}
Using (\ref{S3E15b}) in the formula of $Q\left[  F_{1}\right] $
and plugging the final expression in (\ref{S4E5b}) we obtain,
after some computations:
\begin{align}
&  w(\xi_{1}, \xi_{2}; \eta_{1}, \eta_{2}; t) = 2\delta(\xi_{1}+\xi_{2}-\eta
_{1}-\eta_{2})q[f](\xi_{1}, \xi_{2}; \eta_{1}, \eta_{2}, t) \nonumber %\label{S4E11a} 
\\
& q[f_1](\xi_{1}, \xi_{2}; \eta_{1}, \eta_{2}, t) =
\left[f_1(\eta_{1})\! f_1(\eta_{2})\!
\left(\left(\frac{d}{2\pi \lambda}\right)^3+f_1(\xi_{1})\right)\!
\left(\left(\frac{d}{2\pi \lambda}\right)^3+f_1(\xi_{2})\right)-\right. \nonumber \\
&\hskip 3cm \left.- f_1(\xi_{1})f(\xi_{2})\!
\left(\left(\frac{d}{2\pi \lambda}\right)^3+f_1(\eta_{1})\right)
\left(\left(\frac{d}{2\pi \lambda}\right)^3+f_1(\eta_{2})\right)\right]. \label{S4E11}%
\end{align}
where we have dropped the time dependence of the function $f_1$ in the right hand side of (\ref{S4E11}).
The solution $g_2$ to (\ref{S4E10}) is then:

\begin{align}
g_{2}(\xi_{1}, \xi_{2}; \eta_{1}, \eta_{2}; t) = -\frac{2\, i}{\epsilon}
\delta(\xi_{1}+\xi_{2}-\eta_{1}-\eta_{2})
\int_{0}^{t}e^{-\frac{i}{\epsilon^2} \left[ \varepsilon \left( \xi_{1}\right)+\varepsilon\left( \xi_{2}\right)
-\varepsilon\left( \eta_{1}\right)-\varepsilon\left( \eta_{2}\right)\right] (t-s)}\times \nonumber \\
\times
q[f_1](\xi_{1}, \xi_{2}; \eta_{1}, \eta_{2}; s)ds. \label{S4E12}
\end{align}
where we have used that $g_{2}(\cdot, \cdot; \cdot, \cdot;0)\equiv0$ due to (\ref{S3E4bis1}) and (\ref{S4E4}). The Dirac
measure in (\ref{S4E12}) may be simplified if we define,

\begin{align}
g_{2}(\xi_{1}, \xi_{2}; \eta_{1}, \eta_{2}; t)=\delta(\xi_{1}+\xi_{2}-\eta
_{1}-\eta_{2})
\varphi(\xi_{1}, \xi_{2}; \eta_{1}, \eta_{2}; t), \nonumber %\label{S4E14}
\end{align}
from where, (\ref{S4E12}) gives
\begin{align}
\varphi(\xi_{1}, \xi_{2}; \eta_{1}, \eta_{2}; t) =-\frac{2\, i}{\epsilon}
\int_{0}^{t}e^{-\frac{i}{\epsilon^2} \left[ \varepsilon \left( \xi_{1}\right)+\varepsilon\left( \xi_{2}\right)
-\varepsilon\left( \eta_{1}\right)-\varepsilon\left( \eta_{2}\right)\right] (t-s)}
 q[f_1](\xi_{1}, \xi_{2}; \eta_{1}, \eta_{2}; s)ds \label{S4E15}
\end{align}
Plugging (\ref{S4E15}) in (\ref{S4E9}) and using the symmetry of
$q[f_1]$ with respect to its arguments $\xi_{1}, \xi_{2}, \eta_{1}$
and $\eta_{2}$ we finally obtain the following equation for $f_1$:

\begin{eqnarray}
\partial_{t}f_{1}\left( p_1, t\right) =\frac{4 }{\epsilon^2} \int_0^t
ds\int_{\left(\mathbb{R}^3\right) ^{3}}dp_{2}dp_{3}dp_4\left\{ \cos \left[\frac{1}{\epsilon^2}
(\varepsilon(p_1)+\varepsilon(p_2)-\varepsilon(p_3)-\varepsilon(p_4)) (t-s)\right] \right\} \nonumber \\
\times \delta (p_1+p_2-p_3-p_4)\, q[f_1](p_1, p_2; p_3, p_4; s)
\label{S4E20}
\end{eqnarray}

Non-Markovian Boltzmann equations have been found
in several physical situations  (cf. for example \cite {3}, \cite{ESY}, \cite{19}, 
 \cite{17}, \cite{15} and references therein).
\section{\label{42} The kinetic limit: the Uehling Uhlenbeck  equation.}

\subsection{Derivation of the Uehling Uhlenbeck equation.}

The formal derivation of the U-U equation would then proceed as follows. If we suppose that
\begin{eqnarray}
\label{Deltaenergia}
\frac{1}{\epsilon^2}
(\varepsilon(p)+\varepsilon(p_2)-\varepsilon(p_3)-\varepsilon(p_4)) (\tau-\sigma)>>1
\end{eqnarray}
a simple formal argument gives:
\begin{eqnarray}
\frac{1}{\epsilon^2}\cos \left[\frac{1}{\epsilon^2}
(\varepsilon(p)+\varepsilon(p_2)-\varepsilon(p_3)-\varepsilon(p_4)) (\tau-\sigma)\right]\rightharpoonup \nonumber \\
\pi\, \delta(\tau-\sigma)\delta (\varepsilon(p)+\varepsilon(p_2)-\varepsilon(p_3)-\varepsilon(p_4)) \label{S4E23}
\end{eqnarray}
in the sense of measures, where in that formula $p_2\equiv p_3+p_4-p_1$.
We finally end up with the U-U equation:

\begin{eqnarray}
\partial_{t}f_{1}\left( p_1, t\right) =4 \pi  \int_{\left(\mathbb{R}^3\right) ^{3}}dp_{2}dp_{3}dp_4 \delta (\varepsilon(p_1)+\varepsilon(p_2)-\varepsilon(p_3)-\varepsilon(p_4))\nonumber \\
\times \delta (p_1+p_2-p_3-p_4)\, q[f_1](p_1, p_2; p_3, p_4; t)
\label{S4E23000}
\end{eqnarray}

Notice however  that this approximation requires the condition
(\ref{Deltaenergia}), which, using the original physical variables, can be formulated as the quasiclassical condition $\Delta E\Delta t >> \hbar$.  Equation  (\ref{S4E23000}) is just the equation (\ref{S0E1})-(\ref{S0E3}) written in a different system of units.

\subsection{The loss of validity of the kinetic approximation.}
According to the blow up scenario of  Semikoz \& Tkachev (cf.  \cite{11, 12})  and Pomeau et al. (cf.  \cite{8, 9}) the blow up for the equation (\ref{S4E23000})  takes place in a self similar manner and the distribution of particles has relevant variations in the regions of the space of momentum $p$ whose size rescales like the power $(T-t)^{\beta}$ for some positive $\beta$. In order to describe this region by means of an equation free of parameters we look for self similar solutions of (\ref{S4E23000}). Such solutions have the form  \begin{eqnarray}
&& f(t, p)=(T-t)^{-2\beta -1/2}\,
 \Phi (\xi), \qquad  \xi =\frac{p}{(T-t)^\beta} \label{S4E23004}
 \end{eqnarray}
 where the numerically computed value of $\beta$ is such that $\beta=1.069$  (see \cite {9}). The function $\Phi$, that is of order one, solves then an integro differential equation free of parameters. Notice also that the functional form  (\ref{S4E23004}) tells us immediately the time scales for which the interference effects in  (\ref{S4E20}) cannot be ignored, or more precisely, in dimensional variables, the time scale where $\Delta E\, \Delta t\sim \hbar$. This happens if $
p^2(T-t)\sim \epsilon ^2$ or equivalently if
\begin{eqnarray}
\label{S4E23007}
(T-t)\sim \frac{\epsilon ^2}{p^2}.
\end{eqnarray}
Since $p\sim (T-t)^\beta$ in the self similar region, we obtain that the interference effects appear at times
\begin{eqnarray}
\label{time correlations}
(T-t)\sim \epsilon^{\frac{2}{2\beta+1}}
\end{eqnarray}
For this time scales we have to introduce a boundary layer in order to take the interference effects into account.
\subsection{The correlations become of order one in the boundary layer. }
It turns out that in the same time scale where (\ref{S4E23}) starts failing, the small correlation approximation condition (\ref{S3E4bis0090}) ceases being valid. Indeed, assuming the self similar behaviour (\ref{S4E23004}) we obtain:
\begin{eqnarray}
F_1(\zeta, t)&=&(2 \pi)^3\int_{\mathbb{R}^{3}}f_1(p, t)e^{-i \zeta p}dp\\
&=&(2 \pi)^3(T-t)^{\beta-1/2}\int \Phi(Z)e^{-i \zeta (T-t)^{\beta} Z}dZ\\
&=&(T-t)^{\beta-1/2} \Psi(Z (T-t)^\beta).
\end{eqnarray}
On the other hand, (\ref {S3E16}) and (\ref{S3E17}) yield that for the boundary layer time scale:
\begin{eqnarray*}
G\sim \frac{1}{\epsilon}F_1^3 (T-t)
\end{eqnarray*}
from where, we obtain, using  (\ref{time correlations}):
\begin{eqnarray}
G\sim (T-t)^{2 \beta-1} \sim F_1^2.
\end{eqnarray}
A similar argument shows that $|G_k|\sim \Pi_{m=1}^k |F_1|$ for $k>1$. It then follows that the approximation of the system (\ref{S3E5}) (\ref{S3E6}) by system (\ref{S3E5BIS}) (\ref{S3E6BIS}) breaks down at the time scale (\ref{time correlations}).
\section{\label{6}The boundary layer: analytic description.}

In this Section we derive the set of equations describing the boundary layer where the kinetic approximation is lost. Since, as we have seen, the correlations become of order one in that region, we need to keep a major part of the equations in system (\ref{S3E1}). Using the non-dimensional variables (\ref{S3E600000}), that system becomes

\begin{align}
&  i\partial_{t}F_{k}\left(  x_{1},...,x_{k};y_{1},...,y_{k}, t\right)
=A_{1}+A_{2}+A_{3};\qquad  k=1,2, \cdots  \label{S3E1976} \\
&  A_{1}=- \frac{1}{\epsilon^2}\left(  \sum_{s=1}^{k}\left[  \Delta_{x_{s}%
}-\Delta_{y_{s}} \right]  \right)  F_{k}\left(  x_{1},...,x_{k};y_{1}%
,...,y_{k}; t\right) \label{S3E1976-1}\\
&  A_{2}=\frac{1}{\epsilon}\int_{\Omega}d\xi\left[  \sum_{j=1}^{k}\left[  V\left(  \xi
-x_{j}\right)  -V\left(  \xi-y_{j}\right)  \right]  \right] F_{k+1}\left(  x_{1},...,x_{k},\xi;y_{1},...,y_{k},\xi; t\right) \label{S3E1976-2}\\
&  A_{3}=\frac{1}{2 \epsilon}\left(\frac{d}{\lambda} \right)^3\left[  \sum_{j=1}^{k}\sum_{s=1\;,\;k\neq s}^{k}\left[  V\left(
x_{j}-x_{s}\right)  -V\left(  y_{j}-y_{s}\right)  \right]  \right]
\times\nonumber\\
& \hskip 8.7cm  \times F_{k}\left(  x_{1},...,x_{k};y_{1},...,y_{k}; t\right) \label{S3E1976-3}
\end{align}
The rescaling (\ref{time correlations}) suggest to define new variables as follows:
\begin{align}
&F_k(   x_{1},...,x_{k};y_{1},...,y_{k}; t)=\epsilon^{\frac{2\beta-1}{2\beta+1}\, k}
H_k(  X_{1},...,X_{k};Y_{1}%
,...,Y_{k}; \tau)\\
&T-t=-\epsilon^{\frac{2}{2\beta+1}}\, \tau, \quad x_i=\epsilon^{-\frac{2\beta}{2\beta+1}}\, X_i,\quad  p=\epsilon^{\frac{2 \beta}{2\beta+1}}\, P
\end{align}
Neglecting lower order terms in $\epsilon$ and using that $V(x)=\delta(x)$
we obtain that the functions $H_k$ satisfy at leading order the following system:

\begin{equation}
\label{S3E1978}
\left\{
\begin{array}{l}
i\partial_{\tau}H_{k}\left(  X_{1},...,X_{k};Y_{1},...,Y_{k}, \tau\right)
=A_{1}+A_{2};\qquad k=1,2, \cdots  \\ \
A_{1}=- \left( \displaystyle{\sum_{s=1}^{k}}\left[  \Delta_{X_{s}%
}-\Delta_{Y_{s}} \right]  \right)  H_{k}\left(  X_{1},...,X_{k};Y_{1}%
,...,Y_{k}; \tau\right) \\ \
A_{2}= \displaystyle{\sum_{j=1}^{k}}\left[H_{k+1}\left(  X_{1},...,X_{k}, X_j;Y_{1},...,Y_{k},X_j; \tau\right) \right.\\ \
\hskip 3cm \left.-H_{k+1}\left(  X_{1},...,X_{k}, Y_j;Y_{1},...,Y_{k},Y_j; \tau\right)\right]
\end{array}
\right.
\end{equation}

This system must be solved with the asymptotic condition:
\begin{align}
& H_1(X, Y,  \tau) \sim (-\tau)^{\beta-1/2}\Psi ((X-Y) (-\tau)^{\beta})\quad \hbox{as}\,\,\,\tau \to -\infty \label{asymptoI}\\
& H_k(X_1,\cdots, X_k; Y_1, \cdots, Y_k; \tau)\sim
\sum_{\sigma\in S^{k}}\,\,\prod_{m=1}^{k}\,\,H_{1}\left(
X_{\sigma\left(  m\right)  },Y_{m}; \tau\right) 
\quad \hbox{as}\,\,\,\tau \to -\infty \label{asymptoII}
\end{align}
Notice that formula (\ref{S3E1978}) implies that all the correlation function $G_k$ defined in (\ref{S3E4}) became of order the order of $\Pi_{m=1}^k |F_1|$ in the time scale (\ref{time correlations})
.\\
This problem may be also expressed in the phase space using the Wigner transform that are defined as:
\begin{align*}
\varphi_k(X_1,Ê\cdots, X_k; P_1,\cdots, P_k; \tau)
&=\frac{1}{(2 \pi)^3}\int d\zeta_1\, \cdots\, d\zeta_k\, 
e^{{i\,\sum_{j=1}^k}\zeta_j P_j}\times\\
&\times H_k(X_1-\zeta_1,\cdots,X_k-\zeta_k;
X_1+\zeta_1,\cdots,X_k+\zeta_k; \tau).
\end{align*}
Plugging this into the system (\ref{S3E1978}) we obtain:
\begin{align}
\frac{\partial \varphi_k}{\partial \tau }+&\sum_{j=1}^kP_j\cdot \nabla_{X_j}\varphi_k
=(2 \pi)^{3k}\sum_{j=1}^k\int d\zeta_j d\tilde P_j d\tilde P_{k+1}\,
e^{i \zeta_j(P_j-\tilde P_j)}\times \nonumber\\
&\times\left[\varphi_{k+1}(X_1, \cdots, X_k, X_j-\zeta_j; P_1, \dots, \tilde P_j, \cdots, P_k, \tilde P_{k+1}; \tau )-\right.  \label{S3E1999} \\
&\hskip 0cm 
\left.-\varphi_{k+1}(X_1, \cdots, X_k, X_j+\zeta_j; P_1, \dots, \tilde P_j, \cdots, P_k, \tilde P_{k+1}; \tau )\right];\qquad k=1, 2, \cdots. \nonumber
\end{align}
The asymptotic data as $\tau \to -\infty$ are now determined by:
\begin{align}
&\varphi_1(X; P;\tau)=\varphi_1(P; \tau) \sim (-\tau)^{-\beta -1/2}\Phi\left(\frac{P}{(-\tau )^\beta} \right) \label{S3E2341}\\ \nonumber \\
&\varphi_k(X_1, \cdots, X_k; P_1, \dots, P_k; \tau)
 \sim \frac{1}{(2\pi)^{3k}}
\sum_{\sigma\in S^{k}} \int d\zeta_1\,\cdots d\zeta_k\,\,
e^{i\,\sum_{j=1}^k\zeta_j P_j}\times \nonumber\\ 
&\hskip 3cm \times \prod_{m=1}^{k}
 H_1(X_{\sigma(m)}-\zeta_{\sigma(m)}-X_m-\zeta_m; \tau),\,\,\,
 k>1. \label{S3E2342}
\end{align}
Both systems, (\ref{S3E1978})-(\ref{asymptoII}) and (\ref{S3E1999})-(\ref{S3E2342}) are rather complicated objects to study that we do not consider in detail in this paper. However, the solution of this problem should provide a clear description on how the transition from the kinetic regime to the quantum dominated and highly correlated regime takes place.
\\ 
It is interesting to compute the time for which the correlations appear in physical variables. Using (\ref{time correlations}), we obtain that such scale is given by:
\begin{align*}
T^*-t=\frac{2 m \lambda^2}{\hbar}\left(\frac{a\, \lambda^2}{d^3} \right)^{-\frac{4\, \beta}{2\, \beta+1}}
\end{align*}
where $T^*$ is the time at which  the Uehling Uhlenbeck equation blows up in the original physical units. The range of physical moments that would be described by the systems above (``in the boundary layer'') is
\begin{align*}
p\sim \frac{\hbar}{\lambda}\left(\frac{a\, \lambda^2}{d^3} \right)^
{\frac{2\, \beta}{2\, \beta+1}}.
\end{align*}
Finally the correlation lengths in this boundary layer is
\begin{align*}
x\sim \lambda \left(\frac{d^3}{a\, \lambda^2} \right)^{\frac{2\, \beta}{2\, \beta+1}}.
\end{align*}
\begin{acknowledgments}
The authors thank the hospitality of the Max Planck Institute for
Mathematics in the Sciences where this work has been partially
done. M.E. acknowledges the support from grants UPV00127310-15969
and MTM2005-00714. J.J.L.V has been partially supported by the
Humboldt foundation and the research project MTM2004-05634.\\

\end{acknowledgments}

%\bibliography{apssamp}
%
%
%

\end{document}